\documentclass[aps,prc,twocolumn,showpacs,floatfix,nofootinbib,superscriptaddress,amsmath,amssymb,longbibliography,amsproc]{revtex4-1}
\usepackage{epsfig,dsfont,amssymb,amsmath,amsthm,amsfonts,amsbsy,mathrsfs}
\usepackage{graphicx}
\usepackage{amsmath}
\usepackage{amssymb}%
\usepackage{dcolumn}
\usepackage{multirow}
\usepackage{listings}
\usepackage[dvipsnames]{xcolor}
\usepackage[colorlinks=true,linkcolor=blue,citecolor=blue,urlcolor=blue]{hyperref}
\usepackage{todonotes}

\begin{document}

\title{Structure of odd-mass Ne, Na, and Mg nuclei}

\author{Z.~H.~Sun}

\affiliation{Physics Division, Oak Ridge National Laboratory, Oak
  Ridge, Tennessee 37831, USA}
\affiliation{Department of Physics and Astronomy, Louisiana State University, Baton Rouge, Louisiana 70803, USA}

\author{T.~R.~Dj\"arv}
\affiliation{National Center for Computational Sciences, Oak Ridge National Laboratory, Oak Ridge, TN 37831, USA}
\affiliation{Physics Division, Oak Ridge National Laboratory, Oak
  Ridge, Tennessee 37831, USA}

\author{G.~Hagen}
\affiliation{Physics Division, Oak Ridge National Laboratory, Oak
  Ridge, Tennessee 37831, USA}
\affiliation{Department of Physics and Astronomy, University of
  Tennessee, Knoxville, Tennessee 37996, USA}

\author{G.~R. Jansen}
\affiliation{National Center for Computational Sciences, Oak Ridge National Laboratory, Oak Ridge, TN 37831, USA}
\affiliation{Physics Division, Oak Ridge National Laboratory, Oak Ridge, Tennessee 37831, USA}

\author{T.~Papenbrock}
\affiliation{Department of Physics and Astronomy, University of
  Tennessee, Knoxville, Tennessee 37996, USA}
\affiliation{Physics Division, Oak Ridge National Laboratory, Oak
  Ridge, Tennessee 37831, USA}

\begin{abstract}
The island of inversion is a region of neutron-rich nuclei that are deformed in their ground states. In this region, less is known about the energy levels of odd-mass nuclei, how they evolve with increasing neutron numbers, and how they can be organized into rotational bands. We perform {\it ab initio} coupled-cluster calculations of spectra in odd-mass Ne, Na, and Mg nuclei based on an interaction of chiral effective field theory. 
Our results confirm some tentative spin and parity assignments, predict the structure of nuclei near the neutron dripline, and inform us about rotational bands in this region of the nuclear table. 
\end{abstract}
\maketitle

\section{Introduction}
Neutron-rich nuclei at and beyond the ``magic'' neutron number $N=20$ are deformed~\cite{thibault1975,campi1975,detraz1979,poves1987}; this region is known as the ``island of inversion''~\cite{warburton1990}. Since the discovery of this region, considerable knowledge and understanding has been gained about the even-even nuclei within it~\cite{baumann2007,schwerdtfeger2009,doornenbal2009,doornenbal2016,wimmer2010,ahn2019,crawford2019,tsunoda2020,ahn2022,gray2023,madurga2023}, see, e.g., Ref.~\cite{otsuka2020} for a recent review. However, a search in the NuDat~\cite{nndc} database of nuclear data reveals that much less is understood about odd-mass neon and magnesium nuclei. We do not even know how to sort the measured levels into rotational bands. This is a particular gap in our understanding because such results would tell us how shell structure evolves as neutrons are added. 

Accurately describing these nuclei from the fundamental interaction has been challenging due to the interplay of deformation and continuum effects. The evolution and competition of $sd$ and $pf$ shell physics play a key role in determining the ground state spin of the odd-mass nuclei at $N=20$. The recent observation of $^{28}$O~\cite{kondo2023} 
suggests that this nucleus is not doubly magic and that the island of inversion may extend beyond the two-neutron halo $^{29}$F~\cite{revel2020,Gaudefroy2012,Bagchi2020} into the oxygen isotopes. The experimental study of $^{31}$Ne~\cite{nakamura2009}  
also suggested that the $p-$wave continuum creates a halo in this nucleus, i.e. the unpaired neutron occupies the $p_{3/2}$ instead of the $f_{7/2}$ orbital (which is lower in energy in the conventional shell model).  

In this work, we present {\it ab initio}  computations of odd-mass nuclei in the island of inversion. As we do not include continuum effects, our focus is on the nuclei and states that are sufficiently well-bound and below the neutron separation energy.

We build on the recent {\it ab initio} computations of rotational bands~\cite{caprio2013,wiringa2013,dytrych2013,caprio2015,maris2015,dytrych2020,miyagi2020}, although we follow an approach that is conceptually much simpler. Instead of working with wave functions of good angular momentum, as done in the spherical (no-core or symmetry-adapted) shell models,  we start from symmetry-breaking  mean-field states and perform angular-momentum projections at the end of the computation~\cite{Frosini:2021sxj,hagen2022,sun2024}.  The idea for computing odd-mass nuclei consists of putting the odd nucleon into a single-particle orbital with spin/parity $K^\pi$, following the Nilsson model~\cite{nilsson1955}. Applying angular-momentum projection yields a rotational band with a head that has the nuclear spin/parity $I^\pi=K^\pi$. As the spherical shell model~\cite{mayer1955} guided coupled-cluster computations for doubly-magic nuclei~\cite{hagen2014,hagen2016b,morris2018,hu2022}, the Nilsson model becomes our guide for deformed nuclei. Single-reference methods such as coupled-cluster theory then start from axially symmetric mean-field states and include the short-range (dynamical) correlations that yield the bulk of the binding energy~\cite{sun2024}. The angular-momentum projection then includes long-range (static) correlations and yields the states of the rotational band. For even-even nuclei, this approach has propelled {\it ab initio} computations of deformed nuclei into the mass $A\approx 80$ region~\cite{hu2024,hu2024b}. 

This paper is organized as follows. 
In Sect.~\ref{sec:method} we present the methods, i.e. the Hamiltonian and details about the angular-momentum projection. Section~\ref{sec:ntcl} presents a brief summary of the Nuclear Tensor Contraction Library (NTCL) that allows us to perform {\it ab initio} computations at leadership-class computing facilities. We present our results in Sect.~\ref{sec:results} and a summary in Sect.~\ref{sec:summary}.

\section{Method}
\label{sec:method}
Our single-particle basis consists of states from the spherical harmonic oscillator with spacing $\hbar\omega$ and maximum energy $(N_{\rm max}+3/2)\hbar\omega$.  For the neon, sodium, and magnesium nuclei, we use $N_{\rm max}=8$ and $\hbar\omega=14$~MeV. While this is not sufficient to obtain converged ground-state energies, such model spaces are large enough to accurately capture  rotational bands~\cite{sun2024}.

We want to work in the normal-ordered two-body approximation~\cite{hagen2007a,roth2012} to avoid dealing with residual three-nucleon forces in the coupled-cluster method. This is only valid if the resulting normal-ordered two-body Hamiltonian is a scalar under rotations. However, a deformed reference state breaks rotational invariance of the normal-ordered Hamiltonian,
which means that one cannot simply perform a symmetry-breaking Hartree-Fock calculation and then employ the normal-ordered two-body approximation. Instead, we follow \textcite{Frosini:2021tuj} and perform a spherical Hartree-Fock calculation, where the employed density matrix is a scalar under rotation, by using a fractional filling of the valence shells. We perform the Hartree-Fock computation using such spherical density matrices and this yields a spherical single-particle basis. The Hamiltonian is then normal-ordered and truncated at the two-body level. This completes the first step.

We then use this Hamiltonian, back transformed to the particle vacuum, in the spherical harmonic oscillator basis and employ a symmetry-breaking density matrix that reflects the expected occupation of Nilsson orbitals. This means that we fill pairs of nucleons in time-reversed single-particle states and place the odd nucleon such that its angular momentum projection $J_z$ (taken to be positive) and parity determine the quantum numbers $K^\pi$. We use the laboratory $z$ axis as the symmetry axis of the deformed nucleus. Instead of being guided by the Nilsson diagram, one can also add a mass-quadrupole constraint to the Hamiltonian and map out the Hartree-Fock energy as a function of the quadrupole moment. This is particularly useful for nuclei where neutrons fill the traditional $N=20$ shell. We used both approaches to obtain reference states of interest. These procedures are, of course, also well known from mean-field computations~\cite{bender2003}. 

The result of these procedures is an axially symmetric reference state, with spin/parity $K^\pi$, that can be written as
\begin{equation}
\label{ref}
    |\Phi\rangle \equiv \prod_{i=1}^A \hat{a}_i^\dagger |0\rangle \ .
\end{equation}

Here, $\hat{a}_p^\dagger$ creates a nucleon in the state labelled by $p$, i.e. $|p\rangle = \hat{a}_p^\dagger|0\rangle$ and the vacuum is $|0\rangle$. The single-particle states have good angular-momentum projection $J_z$, good parity, and isospin. Starting with Eq.~\eqref{ref}, we follow the convention that subscripts $i,j,k,\ldots$ label single-particle states occupied in the reference state and $a,b,c,\ldots$ label unoccupied states. We use labels $p,q,r,\ldots$ when no distinction is made.

Coupled-cluster theory~\cite{kuemmel1978,bartlett2007,hagen2014} writes the ground-state as 
\begin{equation}
    |\Psi\rangle = e^{\hat{T}} |\Phi\rangle \ .
\end{equation}
Here, the cluster operator
\begin{equation}
    \hat{T} =\hat{T}_1 + \hat{T}_2 + \hat{T}_3 + \ldots
\end{equation}
consists of one-particle--one-hole (1p-1h), two-particle--two-hole (2p-2h) excitations
\begin{align}
        \hat{T}_1 & \equiv \sum_{ia} t_i^a \hat{a}_a^\dagger \hat{a}_i \ , \\
        \hat{T}_2 & \equiv \frac{1}{4}\sum_{ijab} t_{ij}^{ab} \hat{a}_a^\dagger \hat{a}_b^\dagger \hat{a}_j \hat{a}_i\ ,
\end{align}
and of excitations with higher rank up to and including $A$-particle--$A$-hole ($A$p-$A$h). For most of the paper, we will limit ourselves to including up to 2p-2h excitations. This is the coupled-cluster singles and doubles (CCSD) approximation, and one has to solve a set of nonlinear equations to determine the amplitudes $t_i^a$ and $t_{ij}^{ab}$ for a given Hamiltonian~\cite{shavittbartlett2009}. 

As the reference state is axially symmetric, $|\Psi\rangle $ breaks rotational invariance; it would take up to $A$p-$A$h excitations to restore the symmetry within this formalism, and that is computationally not attractive. However, the CCSD approximation yields about 90\% of the correlation energy and, in particular, includes short-range two-body correlations~\cite{coester1960}. This yields the bulk of the nuclear binding energy~\cite{sun2024}. In contrast, symmetry restoration includes long-range correlations, yields small energy gains for the ground state and reproduces the small spacings within a rotational band~\cite{hagen2022,sun2024}.

For the angular momentum projection, we build on the work by \textcite{qiu2017} and follow Refs.~\cite{bally2018,sun2024}.
As coupled-cluster theory is bi-variational~\cite{arponen1982,arponen1983}, we use the energy functional 
\begin{equation}
E_{J,K} \equiv \frac{\langle \widetilde{\Psi} \vert \hat{P}_{J,K} \hat{H} \vert \Psi \rangle} {\langle \widetilde{\Psi} \vert \hat{P}_{J,K} \vert \Psi \rangle } 
\label{eq:CC_PAV}
\end{equation}
to compute the energy $E_{J,K}$ of the state with total angular momentum $J$ and axial projection $K$. Here, $\langle \widetilde{\Psi} \vert \equiv \langle \Phi \vert (1 + \Lambda)e^{-\hat{T}}$ is the left ground-state, and $\hat{P}_{J,K}$ denotes the  operator 
\begin{equation}
\label{eq:PJ}
    \hat{P}_{J,K} = \frac{2J+1}{2}\int\limits_0^{\pi} {\rm d}\beta    \sin\beta  d^J_{KK}(\beta) \hat{R}(\beta)
\end{equation}
that projects onto angular momentum $J$ with axial projection $K$ (and $z$-axis projection $K$). The rotation operator is 
$\hat{R}(\beta) = e^{-i\beta \hat{J}_y}$, and we employed the Wigner $d^J_{KK}(\beta)$ function~\cite{varshalovich1988}. 
It is convenient to rewrite the energy~(\ref{eq:CC_PAV}) in terms of the Hamiltonian and norm kernels
\begin{equation}
\begin{aligned}
\label{kernels}
    {\cal H}(\beta) &\equiv \langle \widetilde{\Psi} \vert \hat{R}(\beta) \hat{H} \vert \Psi \rangle \ , \\
    {\cal N}(\beta) &\equiv \langle \widetilde{\Psi} \vert \hat{R}(\beta)  \vert \Psi \rangle \ .
\end{aligned}
\end{equation}
as
\begin{equation}
    E_{J,K}=\frac{\int\limits_0^\pi{\rm d}\beta \sin\beta d^J_{KK}(\beta) {\cal H}(\beta)}{\int\limits_0^\pi{\rm d}\beta \sin\beta d^J_{KK}(\beta){\cal N}(\beta)} \ .
\end{equation}

To evaluate the kernels~\eqref{kernels} one inserts the identity $\hat{R}(\beta)\hat{R}^{-1}(\beta)$ and uses Thouless theorem~\cite{thouless1960} to rewrite
\begin{equation}
\label{Vbeta}
    \langle{\Phi} \vert \hat{R}(\beta) = \langle\Phi|\hat{R}(\beta)|\Phi\rangle  \langle{\Phi} \vert e^{-\hat{V}(\beta)}  \ .
\end{equation}
Here $\hat{V}(\beta)$ is a 1p-1h de-excitation operator. After inserting the identity $e^{-\hat{V}}e^{\hat{V}}$ and performing the associated basis transformation, one needs to compute $e^{\hat{V}}e^{\hat{T}}|\Phi\rangle$. It is currently not known how to do that efficiently without approximations. The disentangled method proposed by \textcite{qiu2017} can be used for this purpose, but it does not preserve symmetries in the 
energy and norm kernels~\cite{sun2024}. Instead, we follow Ref.~\cite{bally2018,sun2024} and expand
\begin{equation}\label{lambda_ode}
    e^{\lambda \hat{V}}e^{\hat{T}}=e^{W_0(\lambda)+\hat{W}_1(\lambda)+\hat{W}_2(\lambda)+\ldots} \ .
\end{equation}

Here, the right-hand side consists of the function $W_0$ and the $n$p-$n$h excitation operators $\hat{W}_n$, with $n=1,\ldots,A$. To keep matters computationally tractable, we approximate the right-hand side by only keeping $W_0$, the 1p-1h operators $\hat{W}_1$, and the 2p-2h operators $\hat{W}_2$ while discarding higher-ranking particle-hole excitation operators. Taking the derivative of Eq.~\eqref{lambda_ode} with respect to $\lambda$ yields a differential equation which we solve by integrating from $\lambda=0$ (where $W_0$=0 and $\hat{W}_i=\hat{T}_i$) to $\lambda=1$. 

We note that the application of Thouless theorem~\cite{thouless1960} is limited to non-vanishing vacuum kernels. Thus, it is necessary that the states $|\Phi\rangle$ and $\hat{R}(\beta)|\Phi\rangle$ have a finite overlap $\langle\Phi|\hat{R}(\beta)|\Phi\rangle \neq 0$.  In odd-mass nuclei, this overlap vanishes at $\beta=\pi$ where the unpaired orbital becomes its time-reversed partner. The expression $V(\beta)$ in Eq.~\eqref{Vbeta} then becomes singular. In the vicinity of the point $\beta=\pi$ (and whenever overlaps are becoming exceedingly small in magnitude), we use a singular value decomposition~\cite{robledo2020} when computing a matrix inverse that enters the construction of $V(\beta)$~\cite{qiu2017} and avoid the calculation at $\beta=\pi$.

\begin{figure}[!htbp]
\centering
\includegraphics[width=0.95\linewidth]{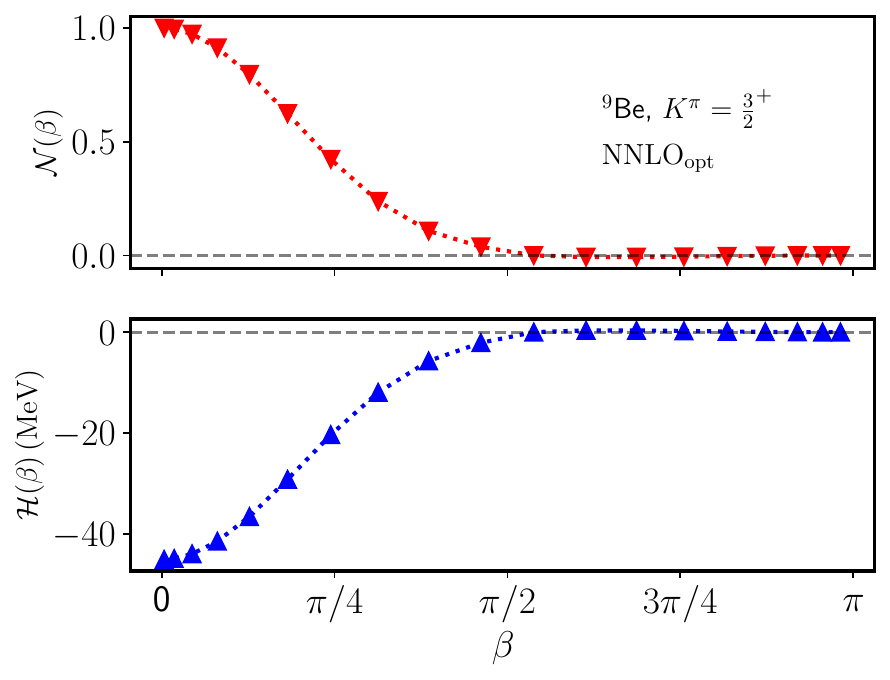}
\caption{Norm and Hamiltonian kernel of $^{9}$Be in $K^\pi=3/2^-$ band, computed with the NNLO$_{\rm opt}$ nucleon-nucleon potential from symmetry-projected coupled-cluster theory. Computations use a model space with $N_{\rm{max}}=8$ and $\hbar\omega = 14$~MeV. }
\label{fig:be9_norm}
\end{figure}

Figure~\ref{fig:be9_norm} shows the norm kernel $\mathcal{N}(\beta)$ and the Hamiltonian kernel $\mathcal{H}(\beta)$ of the $K^\pi=3/2^-$ reference state for the nucleus $^9$Be using the NNLO$_{\rm{opt}}$ interaction~\cite{ekstrom2013}. The kernels fulfill ${\cal N}(2\pi-\beta) = - {\cal N}(\beta)$ and ${\cal H}(2\pi-\beta) = - {\cal H}(\beta)$, and we only show the nontrivial part.  The energy of the unprojected state is ${\cal{H}}(0)$, and we have ${\cal N}(0)=1$. 

\section{Nuclear Tensor Contraction Library}
\label{sec:ntcl}

We calculated the results presented in this work using the NTCL~\cite{ntcl}. This domain-specific, architecture-independent Fortran library runs efficiently at scale on Frontier~\cite{frontier}, the DOE flagship supercomputer located at the Oak Ridge Leadership Computing Facility at Oak Ridge National Laboratory. Frontier is an HPE Cray EX supercomputer with a theoretical peak double-precision performance of approximately two exaflops, consisting of 9408 AMD compute nodes, each with one 64-core AMD ``Optimized 3rd Gen EPYC'' CPU, 512 GB of DDR4 memory, four AMD MI250X GPU’s, and 512GB of high-bandwidth memory (HBM2E) directly on the GPU’s. NTCL makes the hardware on Frontier transparent to the user by presenting a hardware-independent application programming interface to the user, where we have implemented the core computationally expensive operations in hardware-dependent plugins selected when compiling the library.

For the calculations presented in this work, NTCL offloads matrix multiplications to the GPUs by intercepting calls to the \lstinline{*gemm} matrix-multiplication subroutines from the BLAS~\cite{blas} library and replacing them with calls to the rocBLAS library appropriate for Frontier. Since the performance of the projected coupled-cluster code is mainly dependent on efficient tensor contractions that are written as a combination of tensor permutations and matrix multiplication, using NTCL in this way allows us to use Frontier at scale with minimal changes to the projected coupled-cluster code.

To replace BLAS \lstinline{*gemm} calls with rocBLAS \lstinline{*rocBLAS}, NTCL has an interface, \lstinline{ntcl_gemm}, that is designed to have exactly the same signature as BLAS \lstinline{*gemm}. The simplest way to use it is to insert \lstinline{use :: algorithms_api, only : dgemm=>ntcl_gemm} at the top of each Fortran module. NTCL has internal mechanisms to select which matrix-multiplication routines to use and to transfer data from RAM to GPU memory.

NTCL utilizes the factory pattern to decide what routines to use for a given system. Specifically for matrix multiplication, we have written an abstract class \lstinline{matrix_multiplication} that gives a simple-to-use but general interface for matrix multiplication. For each hardware architecture supported by NTCL, we write a separate class that extends \lstinline{matrix_multiplication} as a plugin for that hardware. The correct hardware implementation is then selected by calling a factory class that knows which plugins are available for the system at hand. For example, we have implemented a specific extension of the \lstinline{matrix_multiplication} abstract class that uses rocBLAS and is activated for systems with AMD GPUs. This rocBLAS plugin has been tested and optimized for Frontier.

The matrix data sent to the NTCL-\lstinline{gemm} interface is stored in RAM and needs to be copied to the GPU before the rocBLAS \lstinline{*gemm} routines can be called. NTCL has an internal memory management system that can seamlessly handle heterogeneous memory
architectures, i.e., systems with more than one memory pool, most commonly RAM and GPU memory. This is done by once again utilizing the factory pattern; we have an abstract class representing a general memory pool, we have extensions for each type of memory pool, and a factory class is used to select a specific memory pool. These memory pool classes can then be used to easily transfer data from one type of memory to another, if necessary.

In addition to the NTCL-\lstinline{gemm} interface, NTCL supports general tensor contractions. In this case, NTCL provides a tensor class that represents a general dense tensor that can either be stored in RAM or GPU memory, which allows the program to keep all the tensors in GPU memory throughout the calculation and only copy them back to RAM when the calculations are done. While this functionality is easy to use, significant work would still be required to translate the existing code. The \lstinline{gemm} interface provides a stepping stone that allows you to quickly use GPUs for matrix multiplications, but using the tensor classes to store data in GPU memory is crucial for optimal performance.

\begin{figure}[htpb]
    \centering
    \includegraphics[width=1\linewidth]{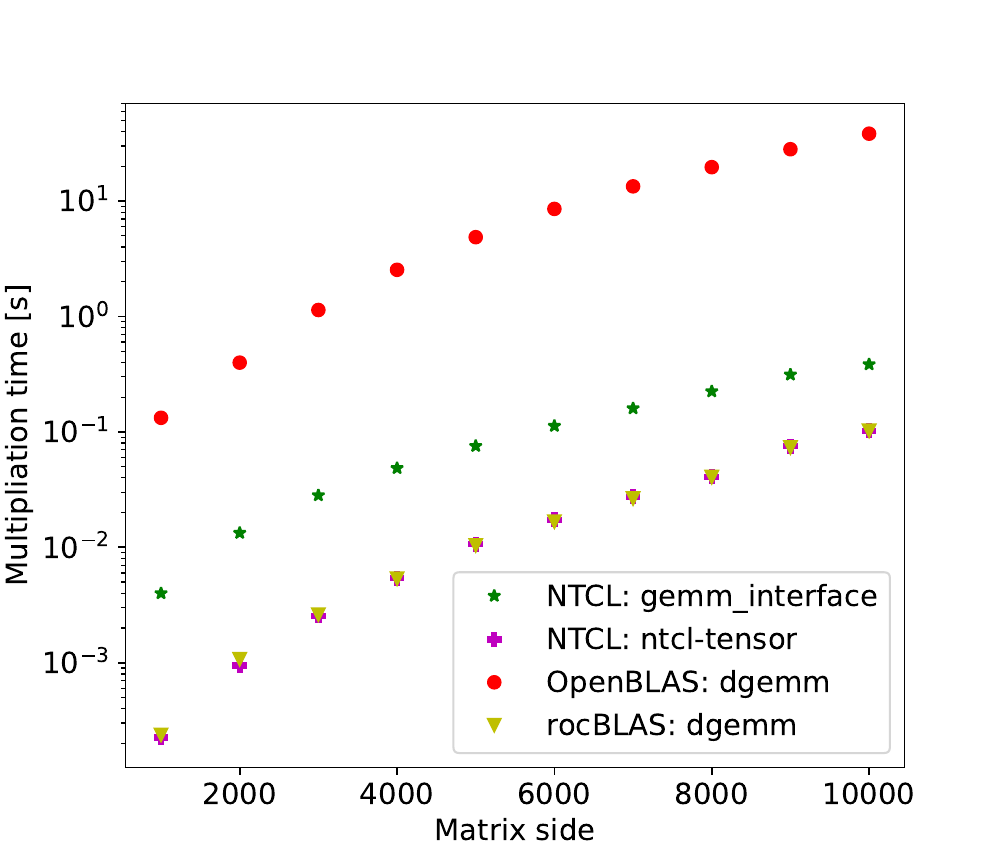}
    \caption{Benchmark of the NTCL for matrix-multiplication, compared to OpenBLAS and rocBLAS. Green stars represent runs with the NTCL-\lstinline{gemm} interface, purple plusses are runs with the NTCL tensor class, red circles are runs with OpenBLAS, and yellow triangles represent runs with rocBLAS.}
    \label{fig:ntcl-bench}
\end{figure}

In Fig.~\ref{fig:ntcl-bench}, we have plotted the execution time of an $n\times n$ matrix-multiplication, performed using the NTCL-\lstinline{gemm} interface (green stars), the NTCL tensor class (purple plusses), OpenBLAS (red circles), and rocBLAS (yellow triangles). While the NTCL-\lstinline{gemm} interface is significantly faster than running the pure CPU OpenBLAS \lstinline{dgemm}, both the NTCL tensor class and rocBLAS versions are even faster still. This is because the matrix data is already in GPU memory before the matrix multiplication occurs in the latter two cases. The execution time of the NTCL-\lstinline{gemm} interface is dominated by data transfer from RAM to GPU memory and back again. However, this benchmark illustrates that even when data is transferred back and forth between GPU memory and RAM, there is still a significant gain over the regular version.

\section{Results}
\label{sec:results}

\subsection{Benchmarks and comparisons}

We start with benchmark computations of rotational bands in $^{9}$Be. To compare with previous no-core shell model computations~\cite{caprio2015}, we use the nucleon-nucleon potential NNLO$_{\rm opt}$~\cite{ekstrom2013}. To quantify success, we take uncertainty estimates from the computations of even neon and magnesium isotopes in Ref.~\cite{sun2024}, where the excitation energies of the $2^+$ and $4^+$ states were assigned uncertainties of 20\% and 15\%, respectively. Thus, computed moments of inertia in even-even nuclei have an uncertainty of about 20\%, and we will use that when judging agreement with data in what follows without showing uncertainty bands. Regarding energy differences between band heads, we will assume that theoretical uncertainties are about 1~MeV. This estimate comes from the energy difference for band-head references computed with $N_{\rm max}=8$ and 12 in Ref.~\cite{sun2024}. 

\begin{figure}[thbp]
\centering
\includegraphics[width=0.95\linewidth]{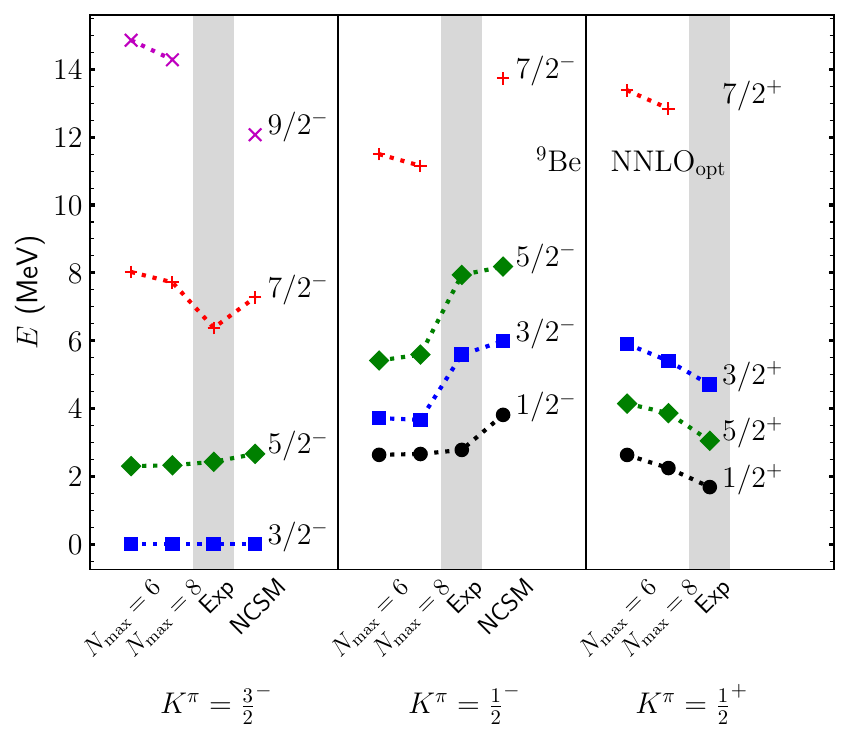}
\caption{Excitation spectra of $^{9}$Be for three bands whose band head has spin/parity $K^\pi$ as indicated, computed with the NNLO$_{\rm opt}$ nucleon-nucleon potential from symmetry projected coupled-cluster theory for $N_{\rm max}=6,8$ and compared to experiment and no-core shell model results from Ref.~\cite{caprio2015}.}
\label{fig:be9}
\end{figure}

Figure~\ref{fig:be9} shows the angular-momentum-projected results from coupled-cluster theory (computed in model spaces with $N_{\rm max} =6$ and 8), compared to the experimental value and to computations using the no-core shell model. 
We see that the $K^\pi=3/2^-$ ground-state band is accurate when compared to data and the no-core shell model benchmark. Here, the reference state is computed by starting with the odd neutron in the $J_z=3/2$ state of the $p_{3/2}$ shell. For the $K=1/2^-$ band, the odd neutron is in the $J_z=1/2$ state of the $p_{1/2}$ shell, and for the $K=1/2^+$ band, the odd neutron is in the $J_z=1/2$ state of the $d_{5/2}$ shell. The head of the $K=1/2^-$ band is at an accurate excitation energy when compared to data, but the the band's moment of inertia is too large. The no-core shell model results are more accurate. It could be that we are in a multi-reference situation for the $K=1/2^-$ band because one could also consider making a hole in the $J_z=1/2$ state of the occupied $p_{3/2}$ shell. 
The current implementation of the projected coupled-cluster method neglects such a potential mixing of different configurations.
Our calculations accurately reproduce the $K=1/2^+$ band in $^9$Be. Overall the coupled-cluster computations are in good-to-fair agreement with benchmarks from the no-core shell model and with data.  

In what follows, we use the interaction 1.8/2.0(EM) of Ref.~\cite{hebeler2011} that is accurate for binding energies and spectra~\cite{hagen2016,hu2022,sun2024}. This interaction consists of nucleon-nucleon and three-nucleon forces. The two-nucleon force is from Ref.~\cite{entem2003}, evolved with the similarity renormalization group~\cite{bogner2007} to a cutoff of 1.8~fm$^{-1}$. The three-nucleon force consists of the leading contributions from chiral effective field theory~\cite{epelbaum2002}. Its cutoff is 2.0~fm$^{-1}$ and the low-energy constants $c_D$ and $c_E$ were adjusted to reproduce properties of nuclei with mass numbers $A=3,4$.

\begin{figure}[htb]
\centering
\includegraphics[width=0.95\linewidth]{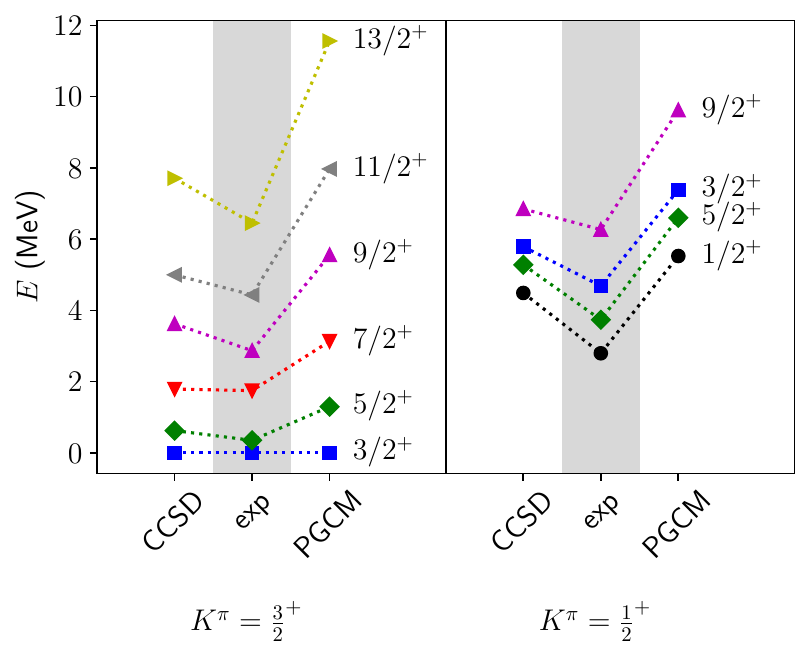}
\caption{Rotational bands in $^{21}$Ne with $K=3/2^+$ and $K=1/2^+$ computed with symmetry-projected coupled cluster of this work (CCSD) and compared to experiment and the projected generator coordinate method (PGCM) results of Ref.~\cite{lin2024}. The employed interaction is 1.8/2.0(EM).}
\label{fig:ne21}
\end{figure}

We start with the nucleus $^{21}$Ne, for which we can compare and contrast our approach to that by \textcite{lin2024}. They used the projected generator coordinate method (PGCM) in an {\it ab initio} setting~\cite{Frosini:2021sxj,bally2021}. Their reference states resulted from Hartree-Fock-Bogoliubov computations of the neighboring even-even nuclei $^{20,22}$Ne. The $^{21}$Ne nucleus was then computed as a quasi-particle excitation of these references. Allowing for different quadrupole deformations of the reference state and projecting onto good particle numbers and angular momentum yielded rotational bands. This approach only captures long-range correlations and does not reproduce binding energies. They found a binding energy of about 119~MeV, compared to our 159~MeV and the experimental 161~MeV. The binding energy and spectrum for $^{21}$Ne remained practically unchanged if $^{20}$Ne or $^{22}$Ne was used as a reference. In Fig.~\ref{fig:ne21}, we compare their results from the $^{20}$Ne reference to ours and to data. Our approach is more accurate when comparing the ground-state band to data and also regarding the excitation energy  of the $K^\pi=1/2^+$ band head. However, the PGCM computation yields a more accurate moment of inertia for the $1/2^+$ band.  We note that the calculations also reveal a $J^\pi=7/2^+$ state in the $K^\pi=1/2^+$ band; the energy is 6.6~MeV for CCSD and 10.1~MeV for the PGCM; however, the data tables~\cite{nndc} did not place a $7/2^+$ level into this band.

\subsection{Results for Na nuclei}

\begin{figure}[htb]
\centering
\includegraphics[width=0.95\linewidth]{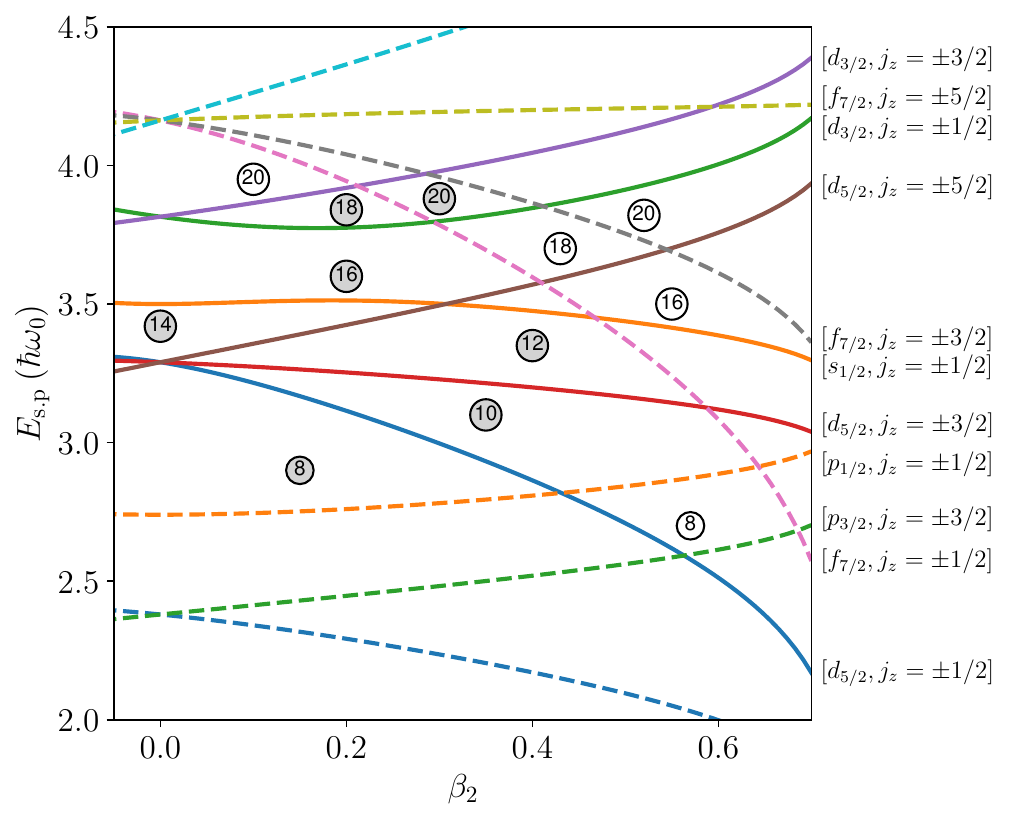}
\caption{Nilsson diagram of single-particle energies as a function of the quadruple deformation parameter $\beta_2$ from Ref.~\cite{BENGTSSON198514}. The dashed (solid) lines represent negative (positive) parity states. Shell gaps are indicated by nucleon numbers and filled circles represent ground-state configurations of the nuclei computed in this work.}
\label{fig:nilsson}
\end{figure}

We computed all sodium isotopes by placing the unpaired proton at [$d_{5/2}, j_z=3/2$] but allowed the neutrons to select different configurations through the self-consistent mean field. 
In a broad range of quadruple deformations ($\beta_2$), the Nilsson diagram suggests a single prolate configuration for protons with $Z=11$, see Fig. ~\ref{fig:nilsson}. 
The $f_{7/2}$ intruder state can become dominant (by forming a $K^\pi=1/2^-$ state) only at larger prolate deformations. 
We assume that all neutrons are paired in time-reversed orbitals and that the calculated odd-mass sodium isotopes have a $K^\pi=3/2^+$ band. Although neutrons do not contribute to the spin and parity of the band head, different configurations of neutrons can yield $K^\pi=3/2^+$ bands with different deformations. In this paper, we focus on the lowest bands with a given spin and parity and neglect any mixing of different deformations.

\begin{figure}[htb]
\centering
\includegraphics[width=0.95\linewidth]{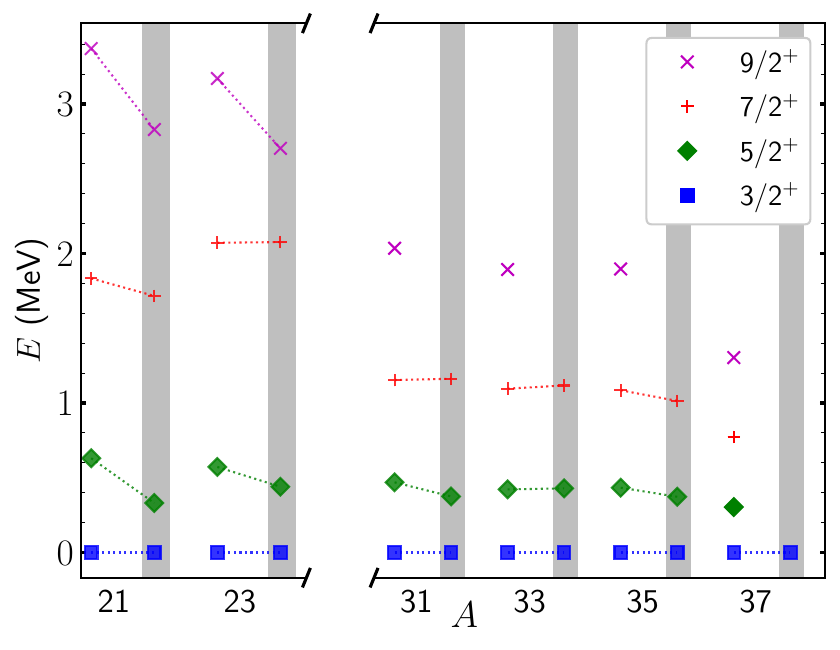}
\caption{Rotational bands with $K^\pi=3/2^+$ of odd-mass Na isotopes with mass number $A$, calculated using the 1.8/2.0(EM) interaction in a model space with $\hbar\omega=14$~MeV and $N_{\rm max}=8$.}
\label{fig:naiso}
\end{figure}

Figure~\ref{fig:naiso} shows the calculation of the ground state bands of sodium isotopes. For neutron numbers $N=10$ and $12$ the Nilsson model suggests that two and four neutrons occupy the $d_{5/2}$ for $^{21}$Na and $^{23}$Na respectively. Our calculations reproduce the $K^\pi = 3/2^+$ bands and agree well with data. We also find that the odd-mass sodium isotopes beyond $N=20$ exhibit a similar rotional structure, consistent with the even-even neighbors~\cite{sun2024}. We predict that $^{37}$Na should have a similar band structure as $^{35}$Na.

In $^{25,27,29}$Na our computations failed to reproduce the near degeneracy of $5/2^+$ and $3/2^+$ states, and instead produced a $K^\pi = 3/2^+$ rotational band similar to those of the  isotopes shown in Fig.~\ref{fig:naiso}.  It seems that $^{25,27,29}$Na exhibit a more complicated structure. First, the data show a $5/2^+$ state that is close to the $3/2^+$ within less than a hundred keV, implying that these nuclei may not be perfect rotors. Second, the low-lying $1/2^+$ states in $^{25,27}$Na suggest either a possible $K^\pi = 1/2^+$ band or that the neutrons are not paired and may yield a non-zero contribution to the spin. 
Finally, the complicated structure of $^{25,27,29}$Na could also be a result of quasi-degenerate neutron configurations  due to the level crossing of $N=14$ and 16, see Figure ~\ref{fig:nilsson}. This possibility is also reflected in the Hartree-Fock calculations of these nuclei. Let us take $^{25}$Na as an example. The normal filling for neutrons is a completely filled $d_{5/2}$ shell, with energies $E_{3/2^+}=-138.5$~MeV and $E_{5/2^+}=-137.9$~MeV from filling the odd proton into different orbitals.  At a larger deformation, the neutron $s_{1/2}$ becomes energetically favorable and a robust local minimum is formed in the energy potential surface, yielding energies $E_{3/2^+}=-139.7$~MeV and $E_{5/2^+}=-139.0$~MeV. Thus, configuration mixing seems possiible. We also investigated the possible excitation of the proton configuration to obtain a $K^\pi = 1/2^+$ band. The resulting energy is $E_{1/2^+}=-138.5$~MeV and $E_{3/2^+}-=137.7$~MeV. The proximity of the energies of three different bands suggests that a multi-configuration approach is called for. Doubting the accuracy of our single-reference approach, we do not show results for $^{25,27,29}$Na. 

We note that other theoretical approaches are also challenged to accurately compute $^{25,27,29}$Na. The shell-model calculations by \citet{sahoo2023} compare results from several contemporary interactions with data and with the highly optimized USD~\cite{warburton1992} interaction. Only the latter is able to accurately describe $^{25,27,29}$Na. Thus, there might also be a deficiency in the interaction we employ.

\subsection{Results for Ne and Mg nuclei}

\begin{figure*}
\centering
\includegraphics[width=0.95\linewidth]{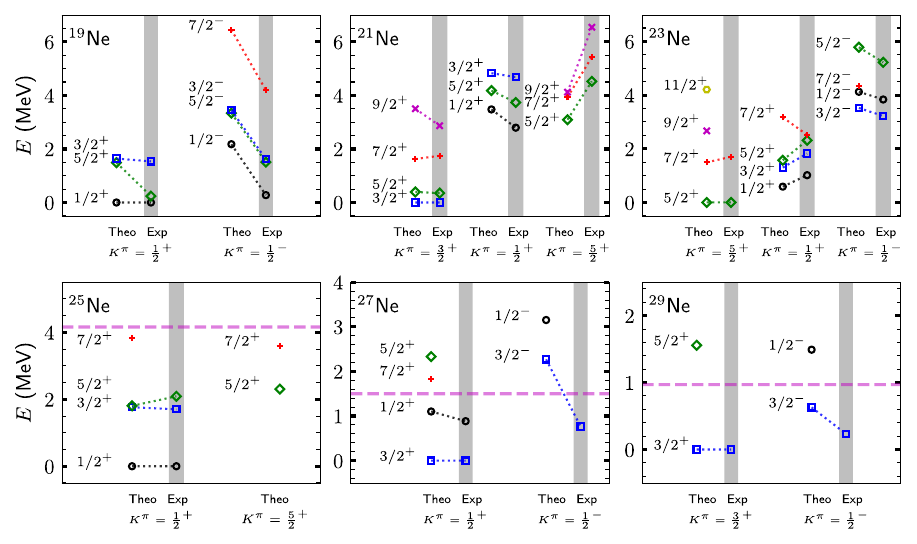}
\caption{Rotational bands of odd-mass Ne isotopes, calculated
using the 1.8/2.0(EM) interaction in model spaces with $\hbar\omega=14$~MeV and $N_{\rm max}=8$. Data, taken from Ref.~\cite{ensdf}, is limited to those states where spin and parities can be assigned by our calculations. The horizontal dashed lines show the experimental neutron separation energy.}
\label{fig:neiso}
\end{figure*}

\begin{figure*}
\centering
\includegraphics[width=0.95\linewidth]{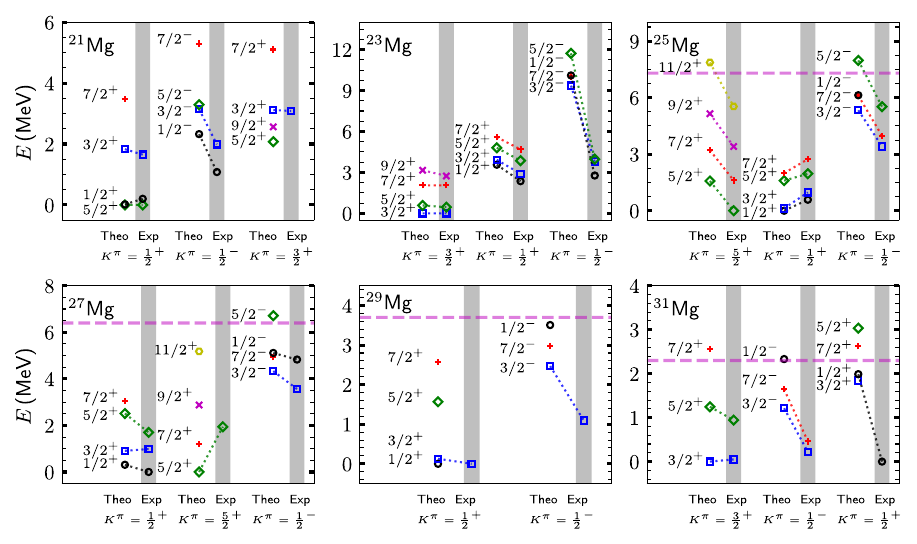}
\caption{Same as Fig.~\ref{fig:neiso} but for Mg isotopes.}
\label{fig:mgiso}
\end{figure*}

Our calculations of the odd-mass neon and magnesium isotopes follow a similar procedure to that of sodium, but the spin and parity of the nuclei are now determined by the configurations of the unpaired neutron. The Nilsson diagram suggests that the protons have a simplified single configuration for $N=10$ and $N=12$ (see Figure ~\ref{fig:nilsson}), i.e. one fills two and four neutrons in the $d_{5/2}$ for neon and magnesium, respectively. Multiple configurations are possible for the neutrons because of intruders from the $pf$ shell. In the even isotopes, the intruder states start to be involved from $N=14$, where a second $0^+$ state gets closer to the ground state and becomes dominant at $N=20$, resulting in the shape coexistence of $^{30}$Ne and $^{32}$Mg~\cite{tsunoda2020,sun2024}. The intruder states may persist until $N=22$ and disappear thereafter, giving nuclei beyond $N=20$ a good single-reference character. One then expects rotationbal bands with similar moments of inertia.

Figures~\ref{fig:neiso} and \ref{fig:mgiso} show our results for the odd-mass neon and magnesium nuclei, respectively. We obtain similar band structures for nuclei with the same number of neutrons. This is expected since the spin and parities of the band heads are determined by the odd neutron.  Our calculation starts at $N=9$ where the last neutron fills the $[d_{5/2},j_z=1/2]$.  This yields the $K^\pi=1/2^+$ bands in $^{19}$Ne and $^{21}$Mg, and our results agree with the data. The excited $K^\pi=1/2^-$ state is obtained by exciting a neutron from $p_{1/2}$ to $d_{5/2}$; this could also be seen as going beyond a level crossing in the Nilsson diagram by increasing the quadruple deformation. We also find a $K^\pi=5/2^+$ band in $^{21}$Mg by exciting the last neutron to the $J_z=5/2$ of the $d_{5/2}$ shell, corresponding to an oblate deformed configuration.

At neutron number $N=11$, one can place the odd neutron in either $[d_{5/2},j_z=3/2]$ or $[d_{5/2},j_z=1/2]$ to get the $K^\pi=3/2^+$ and $K^\pi=1/2^+$ band heads, respectively. Our calculations accurately reproduce the data for $^{21}$Ne and $^{23}$Mg. A negative parity band with $K^\pi=1/2^-$ can be obtained by filling the last neutron in the $[f_{1/2},j_z=1/2]$. However, the resulting band is too high in energy for $^{23}$Mg (and not shown in Fig.~\ref{fig:neiso} for $^{21}$Ne). We see a $K^\pi=5/2^+$ band close to the $K^\pi=1/2^+$ band in $^{21}$Ne.

For the $N=13$ nuclei, we start by filling the odd neutron
in $[d_{5/2},j_z=5/2]$. This yields the $K^\pi=5/2^+$ ground state bands for $^{23}$Ne and $^{25}$Mg and is in agreement with the data. 
According to the Nilsson diagram, the $[s_{1/2},j_z=1/2]$ state could be filled at a larger deformation, leading to the $K^\pi=1/2^+$ bands. Our calculations reproduce the moments of inertia for the $K^\pi=5/2^+$ and $K^\pi=1/2^+$ bands in both nuclei. However, our calculated $K^\pi=5/2^+$ band for $^{25}$Mg is too high in energy. 
For even larger deformations, the Nilsson diagram indicates that the intruder state $[f_{7/2},j_z=1/2]$ is favored in energy, and the angular-momentum projection yields the corresponding negative-parity band. Our calculation is accurate for the moments of inertia, but the band head again is too high in energy for $^{25}$Mg. We note here that the National Nuclear Data Center~\cite{nndc} groups states into bands for $^{25}$Mg.  However, the $1/2^-$ state is assigned to a different band than we suggest in Fig.~\ref{fig:mgiso}. 

As the neutron number is further increased, the neutron separation energies decrease and one slowly approaches the neutron dripline. Separation energies are shown as purple dashed horizontal lines in Figs. ~\ref{fig:neiso} and \ref{fig:mgiso} for neutron numbers $N=15$, 17, and 19.  As our calculations do not use a Gamow basis, the calculations of these nuclei become less reliable. Nevertheless, there is a good-to-fair agreement with data on low-lying states in these nuclei, and one can easily imagine that continuum effects will lower the energy of band heads that are closer to the neutron separation energy.

At neutron number $N=15$ the unpaired neutron is filled in $[s_{1/2},j_z=1/2]$, yielding a $K^\pi=1/2^+$ band in both $^{25}$Ne and $^{27}$Mg. The $K^\pi=5/2^+$ appears as the level crossing of $s_{1/2}$ and $d_{5/2},j_z=\pm 5/2$. The Nilsson model also suggests a level crossing of $[f_{7/2},j_z=\pm 1/2]$ and $[d_{5/2},j_z=\pm 5/2]$, which result in a negative parity band. However, this is not true at the drip line because the nearby continuum favors the $p$-wave (with its small centrifugal barrier) over the $f$-wave. Our calculations do not include continuum effects, and the states near and above the threshold are not accurate. The calculated negative-parity band in $^{25}$Ne is above the threshold. In $^{27}$Mg we find a negative parity band below the threshold and it agrees with data. 

The $N=17$ nuclei mainly have two configurations available, namely the normal $K^\pi=1/2^+$ band by placing the odd neutron in $[d_{3/2},j_z=1/2]$, and the intruder state $f$ orbital yielding a $K^\pi=1/2^-$ band. The data suggest that the $3/2^-$ state is sufficiently bound, and our calculations without continuum are reliable here. We reproduced the $K^\pi=1/2^-$ in both $^{27}$Ne and $^{29}$Mg.

For $N=19$, the Nilsson diagram suggests a series of level crossings between $d_{3/2}$ and $f_{7/2}$. Our calculation reproduces the correct band heads $K^\pi=3/2^+$ and $K^\pi=1/2^-$ in both $^{29}$Ne and $^{31}$Mg. We also find a $K^\pi=1/2^+$ band in $^{31}$Mg, corresponding to the larger deformed $^{30}$Mg plus one neutron at $[d_{3/2},j_z=1/2]$.

\begin{figure}[htb]
\centering
\includegraphics[width=0.95\linewidth]{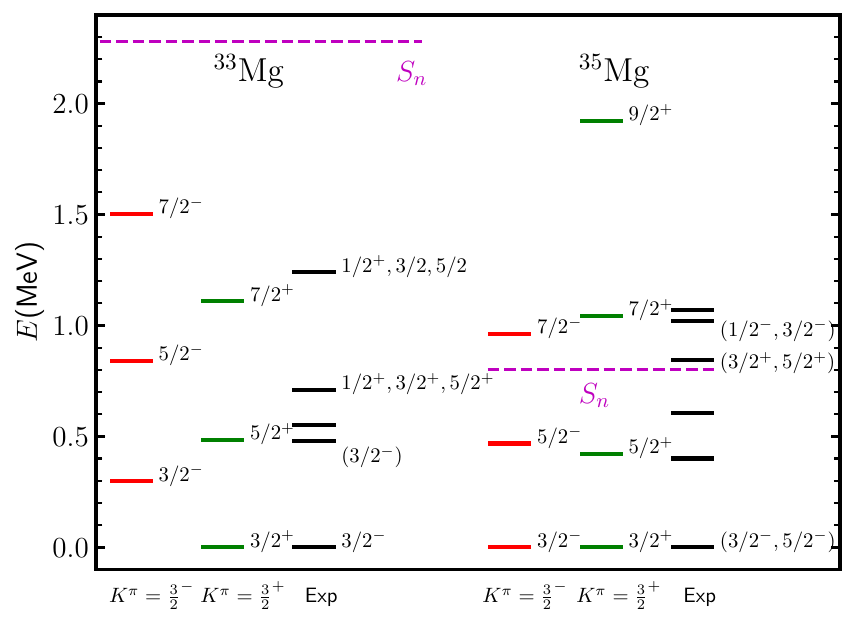}
\caption{Rotational bands with $K^\pi=3/2^\pm$ in $^{33}$Mg and $^{35}$Mg, calculated
using the 1.8/2.0(EM) interaction in model spaces with $N_{\rm max}=8$ and  $\hbar\omega=14$~MeV. Data taken from Ref.~\cite{ensdf}.}
\label{fig:mg3334}
\end{figure}

While neglecting continuum effects has prevented us from making quantitatively accurate predictions for neon and magnesium nuclei beyond $N=20$, studying rotational bands below the threshold would be interesting. Figure~\ref{fig:mg3334} shows our calculated bands for $^{33}$Mg and $^{35}$Mg and compared with the available data.
We obtained the two lowest bands for both nuclei. For $^{33}$Mg, pairs of neutrons occupy the $[f_{7/2},j_z=\pm1/2]$ and $[d_{3/2},j_z=\pm1/2]$ to form a deformed $N$=20 shell. The unpaired neutron can be in $[f_{7/2},j_z=\pm3/2]$ or $[d_{3/2},j_z=\pm3/2]$, yielding the $K^\pi=3/2^-$ and $K^\pi=3/2^+$ bands, respectively.
Our calculations show that both bands exhibit a rigid-rotor pattern. However, our calculations failed to reproduce the correct spin of the ground-state band, presumably because they lack continuum effects.

\section{Summary}
\label{sec:summary}
We presented {\it ab initio} computation of odd mass nuclei in the island of inversion using the projected coupled-cluster theory. We computed the band heads of interest by placing the unpaired nucleon in different single-particle orbits and were guided by  the Nilsson diagram. Our calculation of $^{9}$Be meets benchmarks from the no-core shell model and data. We investigated the low-lying spectrum of stable odd-mass Ne, Na, and Mg nuclei and found overall good agreement with data where band heads can be approximated well by single-reference states. This also allowed us to put states into rotational bands and to predict a few spin/parity assignments. 
 
\acknowledgments
We thank Mark Caprio and Jiangming Yao for sharing their results with us for benchmarking. 
This work was supported by the U.S. Department of Energy, Office of
Science, Office of Nuclear Physics, under Award No.~DE-FG02-96ER40963, by SciDAC-5 (NUCLEI collaboration), and by the Quantum Science Center, a National Quantum Information Science Research Center of the U.S. Department of Energy. Computer time was provided by the Innovative and
Novel Computational Impact on Theory and Experiment (INCITE)
programme. This research used resources of the Oak Ridge Leadership
Computing Facility located at Oak Ridge National Laboratory, which is
supported by the Office of Science of the Department of Energy under
contract No. DE-AC05-00OR22725. 

%
\end{document}